\documentclass{aa}
\usepackage{graphics}
\usepackage{natbib}

\begin{document}

\title{A spatially resolved limb flare on Algol B observed with XMM-Newton}

\author{J. H. M. M. Schmitt, J.-U. Ness, G. Franco}
\institute{
Hamburger Sternwarte, Universit\" at Hamburg, Gojenbergsweg 112,
D-21029 Hamburg, Germany\\
email: jschmitt@hs.uni-hamburg.de}

\date{Received Aug 2, 2003;accepted Aug 22, 2003}
\titlerunning{Algol limb flare}

\abstract{
We report XMM-Newton observations of the eclipsing binary Algol A (B8V) and B
(K2III). The XMM-Newton data cover the phase interval 0.35 -- 0.58, i.e.,
specifically the time of optical secondary minimum, when the X-ray dark B-type
star occults a major fraction of the X-ray bright K-type star. During the
eclipse a flare was observed with complete light curve coverage. The decay part
of the flare can be well described with an exponential decay law allowing a
rectification of the light curve and a reconstruction of the flaring plasma
region. The flare occurred near the limb of Algol B at a height of about 0.1
R$_{\star}$ with plasma densities of a few times 10$^{11}$\,cm$^{-3}$
consistent with spectroscopic density estimates. No eclipse of the quiescent
X-ray emission is observed leading us to the conclusion that the overall coronal
filling factor of Algol B is small.

}

\maketitle

\section{Introduction}

X-ray and radio observations have produced copious evidence for magnetic
field related activity on essentially all cool stars with outer convection
zones \citep{Linsky85,schm97,schmlief,guedelaarev}.
A comparison of solar activity indicators like X-ray luminosity with those
observed for other stars shows that the magnetic activity in the Sun is by
comparison weak. Other stars can have X-ray luminosities exceeding that of
the Sun by up to 4 orders of magnitude, yet this activity is usually
interpreted by analogy with solar activity viewing the emission observed from
stars as arising from scaled-up versions of the corresponding solar features.

Because of the large distance of stars stellar coronae cannot be angularly
resolved and therefore size information is not available. The only exception
comes from eclipsing binaries where size information can be obtained from
light curve analysis. Particularly useful for coronal studies are those
eclipsing binaries where one component can be considered essentially X-ray
dark. Thus one has to look for eclipsing binaries whose primary component is of
spectral type late B or early A. Only very few such systems suitable for X-ray
analysis are known. In the system $\alpha$ CrB 
(\cite{schmkue},\cite{schmitt98},\cite{guedel03}) detected a
total X-ray eclipse during secondary optical minimum, i.e., at the time when
the X-ray dark A star is positioned wide in front of the X-ray bright B star,
allowing them to reduce the size of the corona around $\alpha$ CrB B. Another
such system is the prototypical eclipsing binary Algol, which consists of a
B8V primary and a K2III secondary with an orbital period of 2.87 days. The
relevant system parameters are listed in Tab. 1. Algol is among the apparently
strongest coronal X-ray emitters and has been observed by all major X-ray
satellites. Flares have been reported, for example, by \cite{oord89} and
\cite{white86} who were the first to observe a complete (optical)
secondary minimum; from the absence of any eclipse at secondary minimum, when
the X-ray dark B8V star is in front of the K-type star, they argued for a
corona with large scale height around Algol B. On the other hand, \cite{schmfav}
observed Algol for a whole binary orbit using the Beppo-SAX satellite. They
observed a giant X-ray flare with a decay time of approximately 60 kiloseconds.
At the time of optical secondary minimum a dip in the flare light curve was
observed, which \cite{schmfav} interpreted as an eclipse of the flaring plasma.
From a detailed analysis of the Beppo-SAX X-ray light curve they constrained
the size and location of the flaring region. Interestingly, the flare occurred
above the south polar region of Algol B. Time resolved X-ray spectroscopy could
also be obtained for this flare \citep{favschm}, but from the low-resolution
Beppo-SAX data no spectroscopic density diagnostics was available.

\begin{table}
\caption[ ]{\label{tab1}System Parameters for Algol ($=\beta$ Per)}
\begin{flushleft}
\begin{tabular}{| r | r | r | r |}
\hline
Parameter& Primary & Secondary & System \cr
\hline
Mass (g) & 7.54\ 10$^{33}$ & 1.64\ 10$^{33}$ & \cr
\hline
Radius (cm) & 2.15\ 10$^{11}$ & 2.29\ 10$^{11}$ &\cr
\hline
Spectral type   & B8V       & K2III & \cr
\hline
Luminosity (erg/sec) & 5.95\ 10$^{35}$ & 2.44\ 10$^{34}$ & \cr
\hline
Rotation period (days)   & 2.8673  & 2.8673 & \cr
\hline
Orbital period (days)    &    &  &  2.8673 \cr
\hline
Separation (cm)       &    &  &  1.02\ 10$^{12}$ \cr
\hline
Distance (pc)& & & 28\cr
\hline
Inclination (degrees)& & & 82.5\cr
\hline
\end{tabular}
\end{flushleft}
\end{table}

Here we report an observation of Algol with the EPIC camera and RGS spectrometer
on board XMM-Newton. The observations lasted approximately 45 kiloseconds and
did cover the important phase interval between 0.35 and 0.58, including optical
secondary minimum. The XMM-Newton light curves have much better signal-to-noise
than any previously reported X-ray light curve from Algol and have the specific
advantage of a continuous coverage, while, for example, the Beppo-SAX light
curves were regularly interrupted by earth blocks. Further, the XMM-Newton
observations were simultaneously accompanied by high-resolution spectroscopic
observations with the XMM-Newton RGS, which in particular provide information on
density-sensitive line ratios.

\section{Observations and Data Analysis}

\begin{figure}
\resizebox{\hsize}{!}{\includegraphics{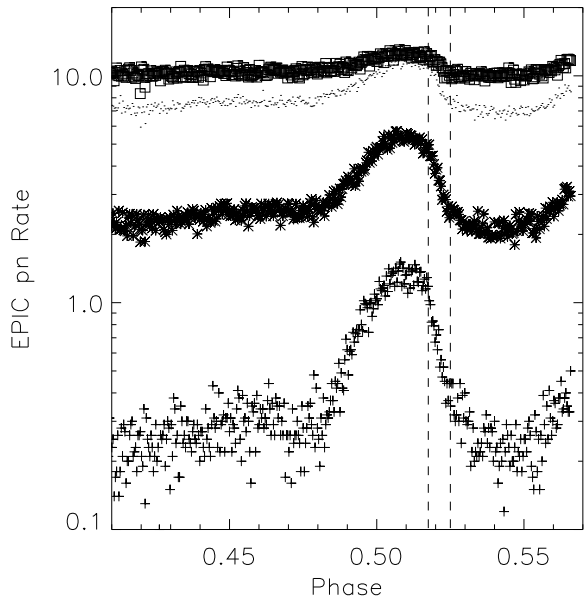}}
\caption[]{\label{f1} EPIC PN light curves of Algol in the energy bands below
1\,keV (squares), between 1--2\,keV (dots), 2--5\,keV (asteriks) and
5--10\,keV (pluses).}
\end{figure}

The Algol system was observed by the XMM-Newton observatory on February
12$^{th}$ 2002. The EPIC-PN detector operated with the thick filter inserted
in the optical path in order to block optical light; data in full-frame window
mode are available
between 04:42:18 and 18:34:35 UTC. We extracted the EPIC PN lightcurve with
standard tools provided by the Science Analysis System (SAS) software version
5.4. Although the source suffers from pile-up due to its substantial
brightness (cf. Fig.~\ref{f1}), we consider this pile-up effect as unimportant
for the timing analysis pursued in this paper. In order to check for pile-up
effects we compared the lightcurves applying different pile-up corrections and
found no significant difference. We also checked the EPIC data for background
proton flares, but none was found in the respective time interval. For our
timing analysis we extracted data from a circular region of 1 arcmin radius
centered on the source while the background data was extracted from a source
free region of the same size on the same CCD chip. A light curve was generated
with the SAS tool {\it evselect}. Inspection of the background light
curve showed it to be totally negligible compared to the observed Algol count
rate; we therefore refrained from any further background subtraction.

In Fig.~\ref{f1} we show the resulting XMM-Newton EPIC light curve vs. phase
in four energy bands, i.e., below 1\,keV, in the band 1--2\,keV, in the band
2--5\,keV, and in the band between 5--10\,keV. The phase refers to the phase
of the primary minimum, which was calculated from the ephemeres used by
\cite{schmfav}. Interestingly, no change in the light curve is seen near the
phase $\phi = 0.43$, i.e., at the time of first contact, when the B type star
starts occulting the K-type star. The light curves start to change at
phase $\phi = 0.48$, when all light curves start increasing, but those at
higher energy in a far more pronounced fashion. The light curves peak near the
phase interval between 0.505 and 0.51, afterwards they start decreasing. At a
phase $\phi = 0.5175$ the decay of all light curves becomes very rapid, and at
$\phi = 0.525$ the light curves in all four energy bands have reached pre-flare
levels. Towards the end of the XMM-Newton observations at a phase of $\phi =
0.555$, the light curves in all energy bands start to increase again.

An inspection of Fig.~\ref{f1} shows that the light curves have the typical
signatures of a stellar flare. There is a tendency for the higher energies
to peak earlier in the flare also the flare spectrum significantly hardens 
compared to the pre-flare emission. For example, the
flux between 5 and 10\,keV increases almost by a factor of 10, while the flux
below 1\,keV increases only by 40\% percent. However, the sharp decrease in the
light curve between phases 0.5175 and 0.525 appears to be very similar in all
energy bands as well as the increase after phase 0.55. It is therefore
extremely suggestive to interpret the dip in the flare light curve between
phases 0.5175 and 0.55 as being caused by an eclipse of the flaring plasma by
the X-ray dark B-type star, in complete analogy to the flare event reported by
Schmitt and Favata (1999). However, in contrast to the Beppo-SAX observations of
Schmitt and Favata (1999) the mid-eclipse was observed at phase $\phi =
0.50$, while the mid-eclipse time of the XMM-Newton flare is at phase $\phi =
0.54$.

\begin{figure}
\resizebox{\hsize}{!}{\includegraphics{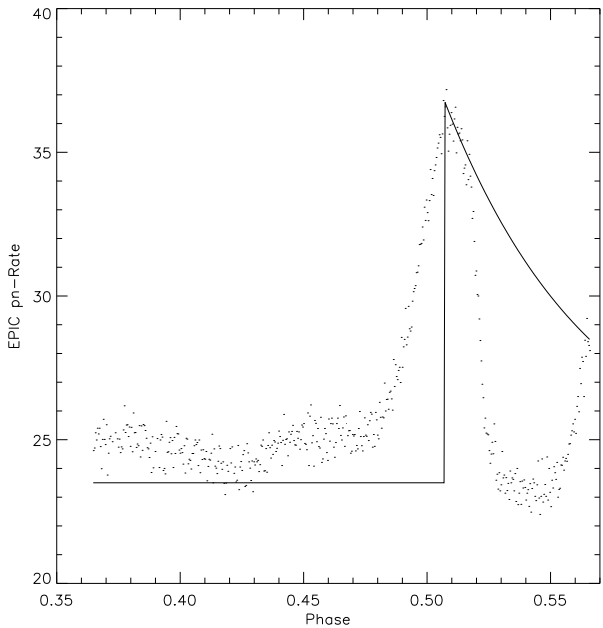}}
\caption[]{\label{f2} EPIC PN light curve of Algol in the total energy band; the solid line
describes a simple exponential decay on top of a constant ``background'' emission; see text
for details.}
\end{figure}

\begin{figure}
\resizebox{\hsize}{!}{\includegraphics{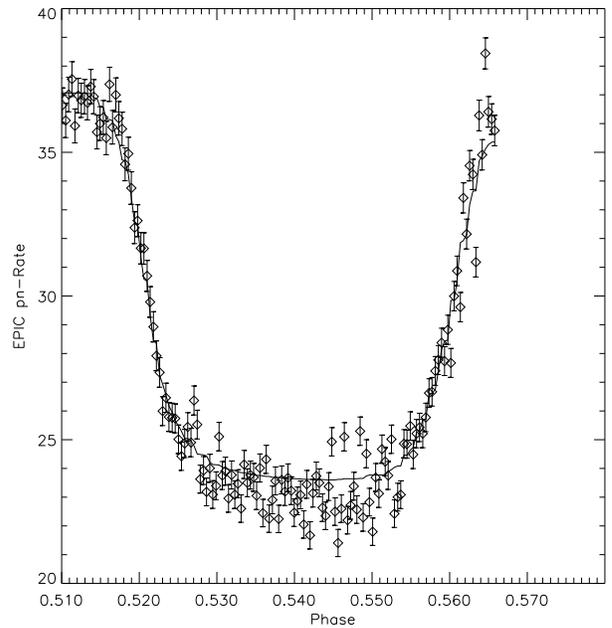}}
\caption[]{\label{f3} Rectified EPIC PN light curve of the flare during its exponential decay
phase; solid line describes the modelled emission (see text for details).}
\end{figure}

In Fig.~\ref{f2} we plot the light curve in the total energy band in a
linear representation.
The solid line shows an exponential decay on top of an assumed constant
``background'' emission, i.e., a light curve with the functional form
$F(t) = A_{0} \ e^{-\frac{t} {\tau}}$, with $A_0$ denoting the
amplitude and $\tau$ the decay time of the flare. It is apparent that the
light curve after the flare peak at phase 0.51 can be well described by a such
simple exponential decay. Exponential decays are very often observed for long
duration solar and stellar flares, and seem to provide a convenient empirical
form to describe flare decay light curves. For this flare on Algol we find
fit parameters of $A_0 = 13.5$\,cts/sec and $\tau = 14.9$\,ksec. Note, that these
values were determined ``by eye'', rather than a formal fit procedure,
and we refrain from giving a formal error analysis.
We emphasize that the light curve data can be
interpreted in such a fashion and that specifically the flux level at the very
end of the observation is consistent with such an interpretation. Using these
values we can rectify the flare decay light curve after the flare peak by
dividing the observed data points by the derived empirical flare decay law.
The resulting light curve is shown in Fig.~\ref{f3}. By construction the light
curve returns to the initial level after the end of the eclipse and the high
symmetry of the flare eclipse light curve becomes apparent.

\section{Reconstruction of the flaring region}

\subsection{Methods}

The late start of the flare eclipse indicates that the flare must have occurred
near the limb of Algol B. From the system geometry it is also clear that the flare
must have occurred on the ``northern'' hemisphere of Algol B, since otherwise the
flare eclipse would have started prior to phase 0.5. In order to reconstruct the
flaring volume we
have to assume that the geometry of the flaring region did not
substantially change during flare progress. Note, that the decrease in
emission measure and hence flux has been taken into account by our
rectification procedure, while changes in the geometry have not been (and
cannot be) accounted for. Clearly, such an assumption is - strictly speaking -
not valid. However, we point out that one may consider -- at least
approximately -- the geometry of many solar flares as constant to first 
order.  Without that assumption no further progress can be made in the stellar
case.  We next introduce a polar
coordinate system with $\theta$ denoting the latitude measured from the north
pole, $\phi _{\rm long}$ the longitudinal angle with the convention that at
orbital phase $\phi_{\rm orb}= 0.5$ the longitudinal angle $\phi_{\rm long} =
0$ is oriented along the central meridian, and $r$ denoting the height above the
stellar surface. Obviously, for a limb flare there is a certain amount of
degeneracy in $\phi_{\rm long}$, since one is looking almost tangentially
through the corona. We therefore treat $\phi_{\rm long}$ as a fixed parameter
which is adjusted to yield the most plausible reconstructed image. To span a
volume we use a grid with $N_1$ (=16) values in $\mu = cos(\theta)$ and $N_2$
(=13) values in $r^3$, so that the whole grid has $N_{\rm tot} = N_1 \times
N_2$ = 208 grid points denoted by some index $j = 1\ldots N_{\rm tot}$. The light
curve is given by $N_{\rm ph}$ data points at phases $\phi_{k}$, denoted by an
index $k=1\ldots N_{\rm ph}$. With the system parameters as specified in
Tab.~\ref{tab1} it is then possible to compute the visibility $V_{j,k}$ of the
volume element $j$ at phase $k$; $V_{j,k} = 1$, if the whole volume element is
visible and contributes to the observed (rectified) count rate CR$_{{\rm obs},k}$
at phase $\phi_{k}$, and $V_{j,k} = 0$, if the whole volume element is
invisible at phase $\phi_{k}$. Let $I_j$ denote the intensity (measured in
cts/sec), produced in the volume element $j = 1\ldots N_{\rm tot}$; clearly, by
construction $I_j$ is constant and refers to the intensity at flare peak.
The intensities $I_j$ are related to the rectified observed count rates
CR$_{{\rm obs},k}$ through the equations
\begin{equation}
{\rm CR}_{{\rm obs},k} = \sum_{j=1}^{N_{\rm tot}} V_{j,k} I{_j} ,\label{c1}
\end{equation}
and we are seeking that set of $I_j$, $j = 1\ldots N_{\rm tot}$, yielding a
``best fit'' to the observed rectified count rates CR$_{{\rm obs},k}$,
$k=1\ldots N_{\rm ph}$, within their error ECR$_{k}$. Obviously, the solution of
Eq.~\ref{c1} represents a ``standard'' astronomical
inversion problem. Specifically, Eq.~\ref{c1} represents a discretized Fredholm
integral equation, and such equations are well known to possess non-unique
solutions. A good overview over inversion problems in astronomy and solution
strategies for such problems is given by \cite{lucy94} and the references given
there. \cite{lucy94} specifically discusses the virtues of regularized and
unregularized solution schemes and points out the importance of the constraint
set by the positivity of the desired solutions. A rather trivial but
nevertheless very important constraint is set by the condition
\begin{equation}
I_{j} \ge 0 .\label{c2}
\end{equation}
One inversion strategy involves minimizing the solution likelihood $H$ defined as
\begin{equation}
H = \sum_{k=1}^{N_{\rm ph}} {\rm CR}_{{\rm exp},k} \ln({\rm CR}_{{\rm obs},k}) ,\label{c3}
\end{equation}
where the quantities CR$_{{\rm exp},k}$, $k=1\ldots N_{\rm ph}$, are the ``expected''
rectified count rates, which are calculated from a given
current solution estimate $I_{j}$, $j = 1\ldots N_{\rm tot}$ through
Eq.~\ref{c1}. This optimization problem
can be iteratively solved by the algorithm \citep{rich72,lucy94}
\begin{equation}
I^{n+1}_{j} = I^{n}_{j}
\frac
{ \sum_{k=1}^{N_{\rm ph}} V_{j,k} \frac {{\rm CR}_{{\rm obs},k}} {{\rm CR}_{{\rm exp},k}} }
{ \sum_{k=1}^{N_{\rm ph}} V_{j,k}}, \label{c4}
\end{equation}
starting with the initial estimate $I^{0}_{j}$ = const. The problem with using
the algorithm in Eq.~\ref{c4} is that there is no obvious criterion when to
stop the iteration. The solution likelihood $H$ continues to decrease on each
iteration, but \cite{lucy94} shows inversions of problems with known solutions
that the best fit reconstructed solutions tend to break up into delta functions,
such that most of the ``flux'' ends up in a rather small number of bins. Thus
the reconstruction of truly continuous distribution functions becomes
rather difficult. \cite{lucy94} shows that stopping the algorithm
after a small number of iterations provides a good solution estimate
and shows that a change in curvature in a $H$-S plot does provide the
derived stopping criterion. Here the entropy S is defined through
\begin{equation}
{\rm S} = - \sum_{j=1}^{N_{\rm tot}} I_{j} \ln(I_{j}). \label{c5}
\end{equation}
\cite{lucy94} argues that such an unregularized reconstruction is superior to
regularized reconstructions because the choice of the value of the parameter
``{$\alpha$}'', which controls the strength of regularization, is arbitrary.
We therefore decided to use Lucy's (1994) inversion scheme.

\subsection{Results}

\begin{figure}
\resizebox{\hsize}{!}{\includegraphics{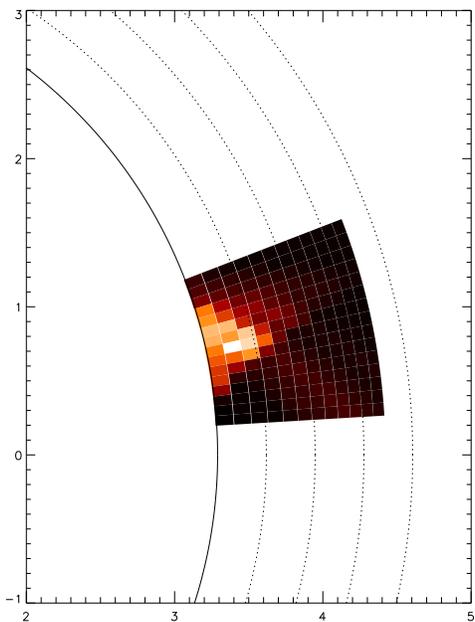}}
\caption[]{\label{f4} Reconstructed spatial distribution of the flare region on Algol B.
All of the flaring plasma has been assumed to be located at a longitude of 
$\lambda = 70^{\circ}$. The circle represents the limb of the 
K-type star, units are in solar
radius; dashed circles indicate heights in steps of 0.1 stellar radii. 
The system is shown at $\phi$ = 0.5556, when the longitude
$\lambda = 70^{\circ}$ is viewed at the limb.}
\end{figure}

Since the shapes of the light curves in different energy bands are very similar,
we performed the reconstruction only for the total band. Next, in order to
fulfill the positivity constraint Eq.~\ref{c2}, we modelled the rectified lightcurve plus
background, with an assumed constant background representing the emission
from the rest of the star. The light curve used for reconstruction as well
as the best fit model light curve resulting from an iteration of Eq.~\ref{c4} is shown in
Fig.~\ref{f3}; the reconstructed flare image is shown in a false-color representation
in Fig.~\ref{f4}. In Fig.~\ref{f4} the solid line shows the apparent limb of the K-type star,
the dotted circles indicate heights in increments of 0.1\,R$_{\star}$.
The important feature of the reconstructed image shown in
Fig.~\ref{f4} and of all other reconstructed images considered by us is that the
flaring region is rather well defined and confined. In particular, at the edge
of the reconstruction volume, the reconstructed intensities become very small, indicating
that the reconstructed size is not determined by the chosen reconstruction volume.

\begin{figure}
\resizebox{\hsize}{!}{\includegraphics{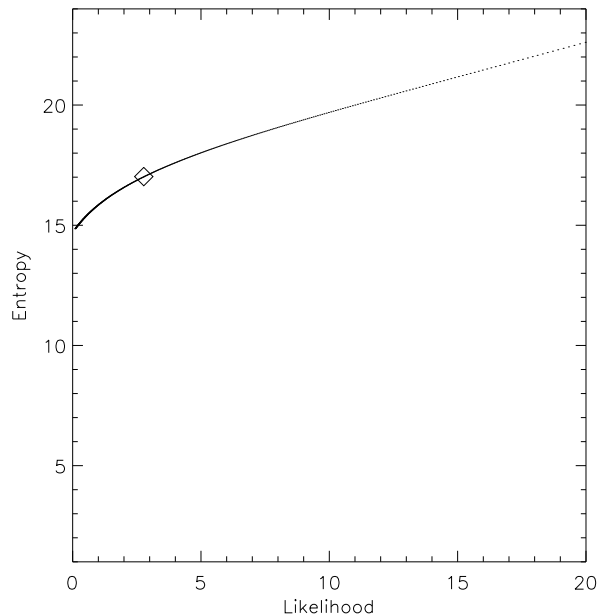}}
\caption[]{\label{f5} Entropy (cf. Eq.~\ref{c5}) vs. likelihood (cf. Eq.~\ref{c3}) for
the iterated solution. The chosen final solution is indicated by a large diamond. }
\end{figure}

As discussed extensively by \cite{lucy94}, it is somewhat difficult to decide
when to stop the iteration in Eq.~\ref{c4}. We followed the recipe
given by \cite{lucy94} and computed for each iteration the likelihood $H$ (cf.
Eq.~\ref{c3}) and solution entropy S (cf. Eq.~\ref{c5}). In Fig.~\ref{f5} we
plot S vs. $H$. There is no clearly identifiable point in the $H$-S diagram
where the curvature changes; the solution chosen by us and shown in
Fig.~\ref{f4} is indicated by a large
diamond. We also investigated the solution properties of further iterated
solutions and found that those pixels with significant flux do not vary substantially
in the course of further iterations, while large flux changes are found only in pixels
whose flux is so small that they do not substantially contribute to the overall emission.

\section{Discussion}

\subsection{Flare energetics}

We first wish to consider the overall energetics of the flare.
For this purpose we need to convert the observed count rate into an energy flux.
In order to convert from count rate to flux we considered various spectra of
active stars (appropriately corrected for photon pile-up) and constructed
acceptable model fits to the EPIC data using a combination of individual
temperature components. We found the energy
conversion factor ECF $= 5.4\ 10^{-12}$\,erg cm$^{-2}$count$^{-1}$
to represent a reasonable choice of count rate to energy flux conversion (in
the 0.2--10\,keV energy band) for active stars. With a distance of 28\,pc
towards Algol, we then compute a peak X-ray luminosity of $6.7 \times 10^{30}$\,erg/sec
for the flare. The total released X-ray energy E$_X$ can then be computed from
the observed decay time of 14.9\,ksec as E$_{\rm tot} \approx 3.6\times
10^{35}$\,erg; note, that this value only refers to the energy contained in the
exponential decay part of the light curve. Clearly, this flare is rather tiny
with only a hundredth of the energy release observed in the giant flare
observed by \cite{schmfav}, yet compared to solar flares, the flare is still
about a thousand times more energetic than the strongest solar flares observed.
As is clear from our reconstructed flare image, the actual size of
the flaring region must have been quite small with a height not exceeding
one tenth of a stellar radius.

\subsection{Size and emission measure of the flaring region}

The longitudinal size of our volume elements is - strictly speaking -
undetermined. However, assuming that the apparent vertical size scale of the
flare agrees with the longitudinal size scale, we deduce a longitudinal size
of $\lambda \sim 6^{\circ}$. Our light curve inversions do yield reasonable
solutions also for other values than $\phi_{\rm long} = 70^{\circ}$, however,
the actual value of $\phi_{\rm long}$ cannot differ too much from that value.
If $\phi_{\rm long}$ was much larger, the K star would self-occult the flare,
and also the rotation of the flare and the apparent motion of the B-type star
would be in opposite directions. If $\phi_{\rm long}$ was much smaller, the
flare eclipse would start too early unless it is located far above the stellar
surface. We therefore adopt - ad hoc - a longitudinal size of $10^{\circ}$ for
each of the volume elements, which then leads to a true size of $V_{\rm elem}
= 1.0 \times 10^{30}$\,cm$^{-3}$ for all the elements. The reconstructed flare
image (cf. Fig.~\ref{c4}) results in an effective count rate in each volume
element. With Algol's distance of d=28\,pc we can then compute the X-ray
luminosity from each volume element through
$L_{X,{\rm elem}}$ = CR$_{\rm elem}$ECF\,4$\pi$d$^2$. The optically thin cooling
function for plasma temperatures at and above 10$^7$\,K is approximately given
by P(T) $\sim 10^{-24.75}$\,T$^{\,0.25}$\,erg\,cm$^3$\,sec$^{-1}$, i.e., it is
rather temperature insensitive. Adopting T $\sim 4.5\times 10^{7}$\,K, which we
determined from spectral fits and also agrees with the values reported by
\cite{yang03}, we can then compute the corresponding emission measure for each
volume element through $EM_{\rm elem} = L_{X,{\rm elem}}/$P(T).

\subsection{Density of the flaring region}

\begin{figure}
\resizebox{\hsize}{!}{\includegraphics{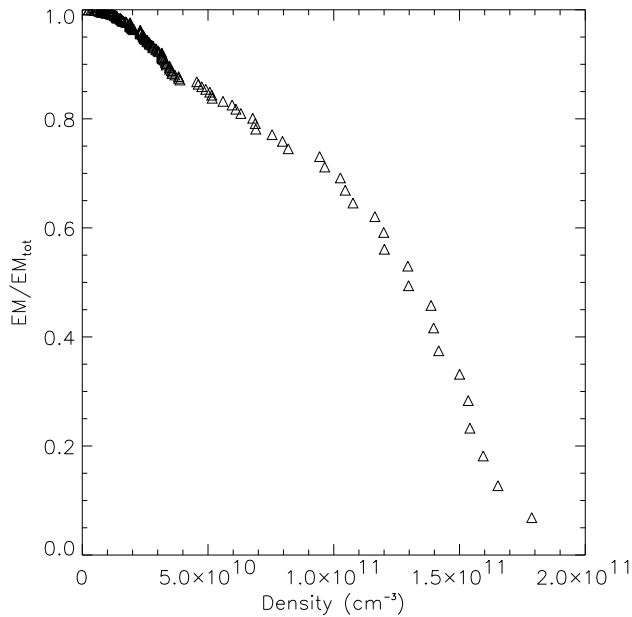}}
\caption[]{\label{f6}Cumulative density distribution histogram for reconstructed
flare image shown in Fig.~\ref{f4}. From the reconstructed intensity in each
volume element and its size, the emission measure and electron density can be
calculated; shown is the density as a function of the integrated and normalized
flare emission measure; see text for details.}
\end{figure}

\begin{figure}
\resizebox{\hsize}{!}{\includegraphics{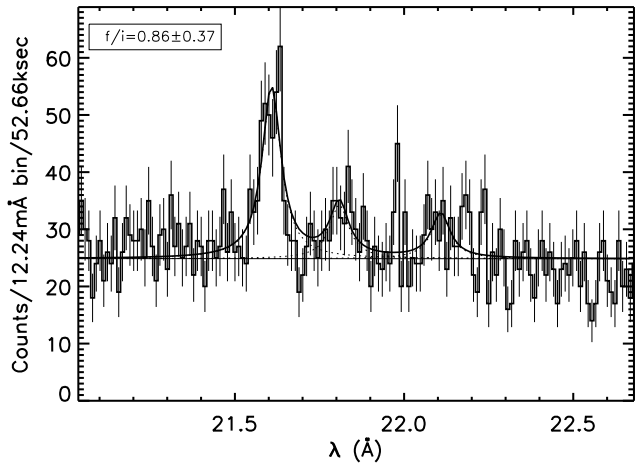}}
\caption[]{\label{f7}O\,{\sc vii} triplet obtained from the full XMM-Newton RGS data set; see text 
for details.}
\end{figure}

Given $V_{\rm elem}$ and $EM_{\rm elem}$ for each volume element, we can
compute the plasma density $n_{\rm elem}$ for each volume
element using $n_{\rm elem} =
\sqrt{EM_{\rm elem}/V_{\rm elem}}$. The resulting density frequency
distribution in our volume elements is shown as a function of the total
integrated and normalized emission measure in Fig.~\ref{f6}. From
Fig.~\ref{f6} it obvious that the bulk of the flare emission comes from
regions with high densities, more than three quarters of the emission
measure resides at densities of at least $10^{11}$\,cm$^{-3}$ and the
peak densities reach approximately 2 $\times$ $10^{11}$\,cm$^{-3}$. With the
simultaneously taken XMM-Newton RGS data (with a slightly longer integration
time of 52.66\,ksec) we can check the densities derived from our light curve
inversions.  Doing this one must be aware that the bulk of the flaring plasma
has a very high temperature (T $\approx$ 45\,MK), where essentially no
O\,{\sc vii} emission is produced, and therefore O\,{\sc vii} emission
actually provides no information on the main flare component.
In Fig.~\ref{f7} we plot the O\,{\sc vii} He-like triplet obtained
from the whole observation. The continuum is very high and the weak intercombination 
and forbidden lines of O\,{\sc vii} can hardly be detected above
the continuum. We nevertheless extracted spectra during quiescent and
flare phases, but we were unable to find significant differences in the f/i
ratios. This indicates that either the sensitivity of our data is such that a
density enhancement between flaring and quiescent emission is not detectable or
that the f/i-ratio is actually dominated by the radiation field of the B-type
star \citep{ness_CS,ness_alg}. Given the geometrical configuration at secondary
minimum, the observable corona on the K-type star is clearly fully immersed in
the B star's radiation field. When accounting for the radiation field in the
way described by \cite{ness_alg} we estimate the density from our measured f/i
ratio f/i = 0.86 $\pm$ 0.37 to $\log\,n_e=10.6\,\pm\,1.42$ (and otherwise
$\log\,n_e=11.1\,\pm\,0.23$).  We also note that the f/i-ratios derived for
Algol from the {\it Chandra} MEG and LETGS and
from the XMM-Newton RGS, i.e., from three independent data sets taken at three
different times, all agree with each other \citep[to within their respective
errors; cf.][]{ness_dens}, so that at present there is no evidence for phase
and/or flare related changes in the O\,{\sc vii} f/i-ratio for Algol.
Unfortunately, with the XMM-Newton RGS it is not possible to disentangle the
contamination due to Fe lines in the Ne\,{\sc ix} triplet \citep[cf.][]{nebr} and
signal in the spectrum obtained from the second dispersion order is too low to
detect the Ne\,{\sc ix} lines.
Since the MEG and HEG f/i ratios derived from {\it Chandra} observations
at different times also do indicate densities in excess of
$10^{11}$\,cm$^{-3}$ \citep[cf.][]{ness_dens}, we conclude that the densities
derived from our light curve inversion are consistent with the available
spectroscopic data on Algol.

\subsection{Cooling of the flare plasma}

An often encountered method to compute densities of a flaring plasma assumes
radiative cooling to be the dominant cooling mechanism. Using the relation
\cite{Doyle88}
\begin{equation}
n_{e,rad} = \frac {3 k T} {\tau {\rm P(T)}},\label{c7}
\end{equation}
where $\tau$ denotes the observed decay time scale of the flare, we compute a density 
$n_{e,rad} \sim 9\times 10^{10}$\,cm$^{-3}$ leading to a
volume of $V_{rad} \sim 2 \times 10^{32}$\,cm$^{3}$, which is of a similar order as
the volume derived in our reconstruction. We therefore conclude that radiative cooling
may in fact be the dominant cooling mechanism for this particular flare observed on Algol B,
in contrast to the flare described by \cite{schmfav}, which cannot have cooled by radiation alone.

\subsection{Magnetic fields}

If we assume the post-flare material to be magnetically confined, we can
compute the minimal field strength $B_{min}$ required for confinement from
\begin{equation}
\frac {B_{min}^2} {8 \pi} = 2 n_e k T\,.\label{c6}
\end{equation}
Using the values $n_e = 2\times 10^{11}$\,cm$^{-3}$ and
$T = 4.5\times 10^{7}$\,K as typical values for density and temperature, we find
$B_{min}\approx $250 G. Assuming that the total released energy ultimately comes
from reconnected magnetic energy, we can compute the typical reconnected, 
i.e., non-potential magnetic field strength from
\begin{equation}
\frac {B^2_{\rm reconnected} V} {8 \pi} = E_{\rm tot} .\label{c8}
\end{equation}
Using $V = 5\times 10^{31}$\,cm$^{3}$, we find an annihilated field of about
430\,G. Of course, since the observed released X-ray energy constitutes only
some (small) fraction of the overall released energy, the latter and hence the
total reconnected magnetic flux must be larger. If we assume - ad hoc - that
only 10\% of the released energy ends up as soft X-ray radiation, reconnected
field strengths $ > 1000$\,G are required. We therefore conclude that coronal
field strengths in excess of 1000\,G are present at least in the lower part of
Algol B's corona. Thus a magnetic reconnection scenario for the Algol flare
appears fully plausible.

\subsection{Quiescent emission}

Finally, it is intersting to note, that at the time of first contact, which
according to the parameters listed in Tab.~\ref{tab1}, is expected to occur at
phase $\phi = 0.43$ no change in the X-ray light curve is apparent.
The light curve stays essentially constant until the flare start at $\phi = 0.48$.
Eclipses of
individual coronal background features might be hidden under the flare-induced
count rate increase, however, the smooth overall appearance of the flare light
curve provides little support for such an assumption. The mid-eclipse count
rates in the energy bands 2--5\,keV and 5--10\,keV agree very well with those
observed prior to flare and eclipse onset, so no eclipses need to be invoked.
In the two lower energy bands the mid-eclipse count rates are reduced by about 10\%
compared to the pre-flare and pre-eclipse rates, on the other hand, the
pre-flare and pre-eclipse rates are also somewhat modulated, so again the
assumption of an eclipse does not appear to the urgently required. We therefore
conclude that the ``quiescent'' emission from Algol B did not experience any
substantial eclipse during our XMM observations and that the vast bulk of the
``quiescent'' coronal emission remained unocculted during the eclipse, in
contrast to the flare which was totally eclipsed.

The absence of any clear eclipse of the ``quiescent'' emission can be explained
either by extended emission or by locating the emission within the
unocculted north polar cap. As pointed out above, all of the available spectroscopic
density measurements of Algol indicate high densities, thus suggesting that
the latter possibility is the far more likely one. With the system parameters
listed in Tab.~\ref{tab1}, we compute that about 10\% of the whole surface
remain uneclipsed during the course of the XMM-Newton observations. Since the
coronal densities of the quiescent emission is very likely \citep[\bf for a detailed
discussion of different density measurements for Algol and other coronal sources
we refer to ][]{ness_dens} not too different
from the density of the flaring region, which we know to be located at a height
of $\approx 0.1$\,R$_{\star}$, we conclude that the ``quiescent'' emission
also comes from rather ``compact'' regions, which must, however, be located at
latitudes $\theta < $ 69$^{\circ}$ from the north pole. The ``southern''
eclipsed part of Algol B must have been devoid of any substantial X-ray
emission at the time of the XMM-Newton observations. Thus the filling factor of
``quiescent'' coronal X-ray emission of Algol B must be small, i.e, $f < 0.1$.
The observed ``quiescent'' count rate of 25\,cts/sec corresponds to an emission
measure of $1.3\times 10^{54}$\,cm$^{-3}$, i.e., only about twice the emission
measure of the flaring region. Assuming that the emission comes from the whole
of the uneclipsed region, we compute for the product $n_e^2 h = 2\times
10^{31}$\,cm$^{-5}$, where $h$ denotes the average height of the corona. If
that average height is the same as that of the flare, i.e, a tenth of a stellar
radius, we compute a density of $3\times 10^{10}$\,cm$^{-3}$. Of course, the
filling factor is likely to be smaller, and if the density is indeed
$10^{11}$\,cm$^{-3}$, the overall filling factor on the K-type star is about
one percent.

\section{Summary}

We have observed a flare on Algol B during secondary minimum with complete
light curve coverage. The location of the flaring region will be well
reconstructed with only some slight ambiguity as to the longitude of the flare
occurrence. In particular, the height of the flare above the surface is about
0.1\,R$_{\star}$ and the flare did not occur in the polar regions. The plasma
densities computed from the observed emission measure and the reconstructed
volume agree with each other and are in the range of a few times
$10^{11}$\,cm$^{-3}$. The ``quiescent'' emission from Algol B was confined to
a region at least 20$^{\circ}$ north of the equator, and only 10\% of Algol B's
surface were not occulted at any time during our XMM-Newton observations.
If quiescent and flaring regions are located at the same heights, minimal densities of
$3\times 10^{10}$\,cm$^{-3}$ result; smaller filling factors result in
correspondingly larger densities as suggested by spectroscopic measurements.
Thus, the overall coronal filling factor of Algol B appears to be quite small
despite its enormous total X-ray luminosity.

\begin{acknowledgements}

We thank Dr. B. Aschenbach (MPE) for communicating to us the work by Yang et al.
prior to submission.  We also thank our referee, Dr. M. G\"udel, for his constructive
criticism which helped to improve our paper.  This work is based on observations
obtained with XMM-Newton, an ESA science mission with instruments and contributions
directly funded by ESA Member States and the USA (NASA).

\end{acknowledgements}

\bibliographystyle{aa}
\bibliography{jn,astron,jhmm}

\begin{thebibliography}{19}
\expandafter\ifx\csname natexlab\endcsname\relax\def\natexlab#1{#1}\fi

\bibitem[{{Doyle} {et~al.}(1988){Doyle}, {Butler}, {Callanan}, {Tagliaferri},
  {de La Reza}, {White}, {Torres}, \& {Quast}}]{Doyle88}
{Doyle}, J.~G., {Butler}, C.~J., {Callanan}, P.~J., {et~al.} 1988, \aap, 191,
  79

\bibitem[{{Favata} \& {Schmitt}(1999)}]{favschm}
{Favata}, F. \& {Schmitt}, J.~H.~M.~M. 1999, \aap, 350, 900

\bibitem[{{G{\" u}del}(2002)}]{guedelaarev}
{G{\" u}del}, M. 2002, \araa, 40, 217

\bibitem[{{G{\" u}del} {et~al.}(2003){G{\" u}del}, {Arzner}, {Audard}, \&
  {Mewe}}]{guedel03}
{G{\" u}del}, M., {Arzner}, K., {Audard}, M., \& {Mewe}, R. 2003, \aap, 403,
  155

\bibitem[{{Linsky}(1985)}]{Linsky85}
{Linsky}, J.~L. 1985, \solphys, 100, 333

\bibitem[{{Lucy}(1994)}]{lucy94}
{Lucy}, L.~B. 1994, Reviews of Modern Astronomy, 7, 31

\bibitem[{{Ness} {et~al.}(2003{\natexlab{a}}){Ness}, {Brickhouse}, {Drake}, \&
  {Huenemoerder}}]{nebr}
{Ness}, J.-U., {Brickhouse}, N.~S., {Drake}, J.~J., \& {Huenemoerder}, D.~P.
  2003{\natexlab{a}}, \apj, in press

\bibitem[{{Ness} {et~al.}(2003{\natexlab{b}}){Ness}, {G\"udel}, {Schmitt},
  {Audard}, \& {Telleschi}}]{ness_dens}
{Ness}, J.-U., {G\"udel}, M., {Schmitt}, J.~H.~M.~M., {Audard}, M., \&
  {Telleschi}, A. 2003{\natexlab{b}}, \aap, in preparation

\bibitem[{{Ness} {et~al.}(2002{\natexlab{a}}){Ness}, {Mewe}, {Schmitt}, \&
  {Raassen}}]{ness_CS}
{Ness}, J.-U., {Mewe}, R., {Schmitt}, J.~H.~M.~M., \& {Raassen}, A.~J.~J.
  2002{\natexlab{a}}, in "The Future of Cool-Star Astrophysics", 2003, Eds. A.
  Brown, G.M. Harper, \& T.R. Ayres Proceedings of 12th Cambridge Workshop on
  Cool Stars, Stellar Systems, \& The Sun, 255--264

\bibitem[{{Ness} {et~al.}(2002{\natexlab{b}}){Ness}, {Schmitt}, {Burwitz},
  {Mewe}, \& {Predehl}}]{ness_alg}
{Ness}, J.-U., {Schmitt}, J.~H.~M.~M., {Burwitz}, V., {Mewe}, R., \& {Predehl},
  P. 2002{\natexlab{b}}, \aap, 387, 1032

\bibitem[{{Richardson}(1972)}]{rich72}
{Richardson}, W.~H. 1972, Optical Society of America Journal, 62, 55

\bibitem[{{Schmitt}(1997)}]{schm97}
{Schmitt}, J.~H.~M.~M. 1997, \aap, 318, 215

\bibitem[{{Schmitt}(1998)}]{schmitt98}
---. 1998, \aap, 333, 199

\bibitem[{{Schmitt} \& {Favata}(1999)}]{schmfav}
{Schmitt}, J.~H.~M.~M. \& {Favata}, F. 1999, \nat, 401, 44

\bibitem[{{Schmitt} \& {K\"urster}(1993)}]{schmkue}
{Schmitt}, J.~H.~M.~M. \& {K\"urster}, M. 1993, Science, 262, 215

\bibitem[{{Schmitt} \& {Liefke}(2002)}]{schmlief}
{Schmitt}, J.~H.~M.~M. \& {Liefke}, C. 2002, \aap, 382, L9

\bibitem[{{van den Oord} \& {Mewe}(1989)}]{oord89}
{van den Oord}, G.~H.~J. \& {Mewe}, R. 1989, \aap, 213, 245

\bibitem[{{White} {et~al.}(1986){White}, {Culhane}, {Parmar}, {Kellett},
  {Kahn}, {van den Oord}, \& {Kuijpers}}]{white86}
{White}, N.~E., {Culhane}, J.~L., {Parmar}, A.~N., {et~al.} 1986, \apj, 301,
  262

\bibitem[{{Yang} {et~al.}(2003){Yang}, {Lu}, {Aschenbach}, {Chen}, {Chen},
  {Jia}, \& {Chen}}]{yang03}
{Yang}, X., {Lu}, F.~J., {Aschenbach}, B., {et~al.} 2003, \aa, submitted

\end{thebibliography}

\end{document}